\documentclass[pdflatex,sn-apa]{sn-jnl}

\usepackage{graphicx}%
\usepackage{multirow}%
\usepackage{amsmath,amssymb,amsfonts}%
\usepackage{amsthm}%
\usepackage{mathrsfs}%
\usepackage[title]{appendix}%
\usepackage{xcolor}%
\usepackage{textcomp}%
\usepackage{manyfoot}%
\usepackage{booktabs}%
\usepackage{algorithm}%
\usepackage{algorithmicx}%
\usepackage{algpseudocode}%
\usepackage{listings}%
\usepackage{soul}%
\usepackage{array}
\usepackage{makecell}

\begin{document}

\title[Article Title]{4D Synchrotron X-Ray Multi Projection Imaging (XMPI) for studying multiphase flow dynamics and flow instabilities in porous networks}


\author*[1,2]{\fnm{Patrick} \sur{Wegele}}\email{Patrick.Wegele@voith.com}

\author*[3]{\fnm{Zisheng} \sur{Yao}}\email{zisheng.yao@fysik.lu.se}

\author[1]{\fnm{Jonas} \sur{Tejbo}}

\author[3]{\fnm{Julia K.} \sur{Rogalinski}}

\author[3]{\fnm{Zhe} \sur{Hu}}

\author[3]{\fnm{Yuhe} \sur{Zhang}}

\author[1]{\fnm{Erfan} \sur{Oliaei}}

\author[1]{\fnm{Saeed} \sur{Davoodi}}

\author[4]{\fnm{Alexander} \sur{Groetsch}}

\author[5]{\fnm{Kim} \sur{Nyg\aa rd}}

\author[3]{\fnm{Eleni Myrto} \sur{Asimakopoulou}}

\author[1]{\fnm{Tomas} \sur{Rosén}}

\author[3]{\fnm{Pablo} \sur{Villanueva-Perez}}

\author*[1]{\fnm{L. Daniel} \sur{Söderberg}}\email{dansod@kth.se}

\affil*[1]{\orgdiv{Department of Fibre and Polymer Technology}, \orgname{KTH Royal Institute of Technology}, \orgaddress{\street{Teknikringen 56}, \city{Stockholm}, \postcode{SE-10044}, \country{Sweden}}}

\affil[2]{\orgname{J.M. Voith SE \& Co. KG}, \orgaddress{\street{St. Poeltener Strasse 43}, \city{Heidenheim an der Brenz}, \postcode{89522}, \country{Germany}}}

\affil[3]{\orgdiv{Synchrotron Radiation Research and NanoLund}, \orgname{Lund University}, \city{Lund}, \country{Sweden}}

\affil[4]{\orgdiv{Department of Engineering Mechanics}, \orgname{KTH Royal Institute of Technology}, \city{Stockholm}, \country{Sweden}}

\affil[5]{\orgdiv{MAX IV Laboratory}, \orgname{Lund University}, \city{Lund}, \country{Sweden}}

\abstract{Flow instabilities such as Haines jumps in porous media are common phenomena that occur on sub-second timescales. X-rays are particularly suitable for investigating these processes because they provide non-destructive three-dimensional insight into the network structure and the liquid distribution within porous media. Studying imbibition events in four dimensions (three spatial dimensions plus time) is inherently challenging with conventional tomography because the required rapid sample rotation imposes significant centrifugal forces that alter the flow. Here, we demonstrate synchrotron X-ray multi-projection imaging~(XMPI) to capture four-dimensional flow in an additively manufactured, homogeneous spherical pore network at 1.3~µm effective pixel size and 50~Hz temporal resolution without the need for high rotational speeds. This enables in situ visualization of non-repeatable pore-scale events in both space and time, a capability unachievable with classical X-ray tomographic approaches. We compare the results to Shan–Chen multiphase Lattice Boltzmann simulations performed on the same geometry, finding both qualitative agreements and systematic differences in filling sequences and timescales. These discrepancies expose key limitations of current simulation methods in representing contact-line dynamics and realistic boundary conditions limitations that XMPI can directly overcome. By enabling high-resolution, real-time imaging of flow instabilities in opaque porous media, synchrotron XMPI provides a unique platform that bridges the gap between pore-scale experiments and simulations.}


\maketitle


\section{Introduction}
Liquid movement in porous media is a common phenomenon in numerous fields, such as fuel cells \citep{Sinha2007}, oil recovery \citep{Blunt1998}, composites and carbon storage \citep{Blunt2013, Zacharoudiou2018, FARAMARZI2021}. When one fluid displaces another within a porous medium, the process is conventionally categorized as either imbibition or drainage, where, in the case of imbibition, a wetting fluid displaces the nonwetting fluid out of the voids in the solid matrix. To describe the capillary imbibition behaviour in macroscopic networks, the Lucas-Washburn equation is a well-established method, assuming that the porous network is formed as a collection of capillary tubes. Consequently, when investigating the imbibition mechanisms on a pore level, this simplification is no more valid. Furthermore, the detailed pore and corner geometry has been identified to be of great importance for the imbibition dynamics \citep{Lenormand1984, Senden2000, HUGHES2000}. On top of that, the microscopic pore filling may occur in the scale of a millisecond while the macroscopic liquid imbibition can range to several hours \citep{Dubé2007}. Therefore, other concepts are applied to describe the pore filling on the microscale \citep{Lenorman1983, Berg2013}.

Fluid transport on the pore scale reflects an interplay between the geometry of the pore network, surface chemistry and rheological properties of the fluids involved. These properties are typically quantified by the static solid-liquid contact angle $\theta_{sl}$, the dynamic viscosity $\mu$ and the surface tension $\sigma$. The relative magnitude of viscous to capillary forces is therefore quantified by the capillary number, $Ca =\mu v / \sigma$, which uses the flow velocity~$v$ to provide a dimensionless measure for classifying imbibition and drainage regimes. Fundamental work was carried out by \cite{Lenorman1983}, describing the principal displacement process of a wetting fluid entering a two-dimensional channel network. Specifically referring to the imbibition behaviour, they showed that imbibition is governed by the number and spacing of the geometries filled with the nonwetting fluid. 

One key mechanism during the liquid imbibition and drainage of porous structures is the so-called \textit{Haines jump} \citep{Haines1930}, which describes an instantaneous filling of pores once a threshold pressure $p_{th}$ created by a pore connecting throat is overcome \citep{Wilkinson1983}. When approaching the throat, the fluid meniscus remains stationary until the local capillary (or externally applied) pressure surpasses this threshold pressure, whereupon the interface bursts through the constriction and instantaneously fills the adjoining pore. The stored energy in the meniscus of the fluid front is released instantaneously, followed by an increase in the kinetic energy of the liquid during the filling of the pore. This snap-through produces a characteristic, abrupt drop in capillary pressure \citep{Berg2013, Andrew2015} and initiates rapid fluid redistribution in the surrounding pore network, which can span up to 20 neighboring pores \citep{Bultreys2024}. \cite{Moebius2012} showed, that the interfacial velocity during the jump can exceed 50~times the average fluid front velocity, causing inertial effects to be relevant during this procedure. Recent work highlights the influence of the contact angle of the wetting liquid on the burst pressure \citep{Yan2025} or provides methods to estimate the relationship between total-, viscous- and capillary pressure drop during Haines jumps \citep{Guo2025}.

Haines jumps can occur during imbibition and drainage processes. In case of forced imbibition, however, the pore-scale displacement mechanisms are inherently different and more complex than in drainage \citep{Joekar-Niasar2012, Lenormand1990}, meaning that imbibition cannot be treated as a simple reversal of drainage processes \citep{Liu2022}. Nevertheless, this asymmetry remains relatively underexplored in experimental studies compared to drainage experiments. This was also particularly confirmed for the Haines jumps, where instabilities during imbibition occur at lower rates \citep{DiCarlo2003}.

While investigations show that the timescale of the Haines jumps depends on the fluid properties and the local pore and throat geometries in the jump zone \citep{Mohanty1987}, ranges of milliseconds are common for pores in the microscale \citep{DiCarlo2003, Mohanty1987, Armstrong2013}, creating additional challenges when experimentally investigating Haines jumps. Therefore, common experimental methods to capture such flow instabilities in pore filling have included acoustic methods \citep{DiCarlo2003}, optical and highspeed imaging \citep{Lenorman1983, Yiotis2021, Moebius2012, Armstrong2013, Saad2022}, confocal microscopes \citep{doNascimento2019, Odier2017, Yan2025} and time-resolved X-ray computed tomography measurements \citep{Berg2013, Bultreys2016, Andrew2015, Singh2017, Meisenheimer2020, Bultreys2024}. All mentioned approaches have significant drawbacks when it comes to resolving the full 3D~structure during imbibition \citep{Lenorman1983, Moebius2012, Saad2022, Yiotis2021}, the dynamics of the sub-second imbibition processes \citep{Berg2013, Bultreys2016} or investigating opaque systems \citep{Odier2017, doNascimento2019}, which are common in the field of porous media. Besides that, numerical approaches exist that are used to investigate the imbibition dynamics of porous networks. Among those, Lattice-Boltzmann-based simulations are the main methods to be used (see \cite{Liu2016} for a comprehensive review). To simulate the multiphase flow in porous media, the most predominant diffuse interface models are the Shan-Chen Model \citep{Shan1993} used for example by \cite{Pan2004, Yang2025} or the color Lattice-Boltzmann model \citep{Rothman1988, Gunsten1991}, used for example by \cite{Yamabe2015, Liu2022, Zhao2024}. 

To overcome existing experimental limitations, time-resolved X-ray computed tomography is a promising method, even though the temporal resolution is limited by the need to rotate the sample without significant liquid movement. Consequently, realistic temporal resolutions are in the range of $10^{-1}-10^{-2}$~Hz, which is insufficient to investigate Haines jump events in the millisecond range. Recently, an investigation of the jump dynamics was possible by either using high viscous fluids, which reduces the filling speed during the Haines jump \citep{Bultreys2024} or assuming a reversible imbibition and drainage behaviour and reconstructing the Haines jump dynamics from several individual events that are observed from different angles \citep{Tekseth2024}. However, no method is yet available that allows for resolving Haines jumps of low-viscosity fluids in 4D~during imbibition processes without assuming reversible imbibition-drainage behaviour.

To address these methodological limitations, we present a study employing the novel synchrotron X-ray multi-projection imaging (XMPI) technique to investigate non-repeatable multiphase flow and imbibition in porous networks in~4D. This approach enables the resolution of fluid dynamics within opaque porous structures at an effective pixel size of 1.3~µm and a temporal resolution of up to 10~kHz~\citep{rogalinski2025timeresolved}. Conventional high-frame-rate tomography would require rapid sample rotation to collect a sufficient number of X-ray projections for 3D~reconstruction, generating significant centrifugal forces that hinder realistic flow experiments. By contrast, the XMPI method overcomes this constraint by using several angularly resolved beamlets that illuminate the sample simultaneously. Only a slight rotation of the sample at low rotational speeds is needed to minimize the difficulty of 4D reconstruction, not imposing significant centrifugal forces while still providing the high temporal and spatial resolution needed to capture pore-scale multiphase flow in situ. Furthermore, it does not require a repeatability of the imbibition process, thereby allowing for the study of more regular and spherical networks or networks that deform during imbibition due to swelling.

In this study, we aim to visualize Haines-jump events in~4D during imbibition into a homogeneous spherical porous medium. This is of particular importance, as the jumping behaviour happens on fast timescales and depends on the pore-throat aspect ratio, which can be precisely controlled within our setup. As there is clear evidence that the dynamics of the pore filling influence the overall formation of a sharp or disperse fluid front on the macroscale \citep{Armstrong2013}, there is a clear need to investigate this process in more detail.

\section{Materials}
\label{p6:Sec_Materials}
\begin{figure}[htbp]
    \centering
    \includegraphics[width=0.5\textwidth]{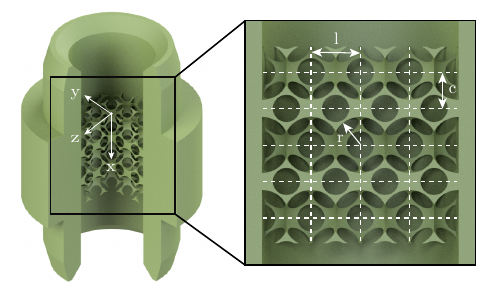}
    \caption{Cut through the 3D~model of the capillary that was used during the experiment. The hollow-spherical structure is visible in the middle, while a circular section above and below is used to provide in- and outflow areas}
    \label{p6:fig_Capillary}
\end{figure}

Within our experiment, we used a custom-designed porous network as depicted in Fig.~\ref{p6:fig_Capillary}, which was additively manufactured from a methacrylate-based resin (Nanoscribe IP-Q). The porous network itself consists of hollow spheres of radius $r=0.100$~mm with a lateral center distance of $l=0.175$~mm and a vertical center distance of $v=0.130$~mm. Six layers of hollow spheres were used to obtain a resulting network height of $1.05$~mm. As both $l>2r{}$ and $v>2r$, no closed pores exist and all pores are accessible to fluid flow. While the pore size of the central pores is $r$, the pores are laterally connected by throats of $r_l=0.049$~mm and vertically connected by throats of $r_v=0.044$~mm. The whole network is situated in a circular duct of diameter $2R=0.7$~mm, which was connected to a Kapton tube for supply with deionized~(DI) water. 

\section{Methods}
\label{p6:Sec_Methods}
In the experiment, we expected the following forced imbibition behaviour: Water as the wetting fluid advances through the microcapillary under a pressure-driven flow, forming a meniscus that contacts the solid walls with the advancing contact angle~$\theta_{a}$. When the meniscus encounters a pore throat with radius~$r_l$ or $r_v$, the narrow constriction generates a capillary entry barrier, preventing the front from entering it. Once the applied driving pressure~$\Delta p$ satisfies the condition~$\Delta p>p_{th}$, the meniscus rapidly invades the adjacent pore in a sudden event known as a Haines jump. After each jump, the next throat in the network becomes the new capillary barrier, resulting in a pore-by-pore progression of the liquid front through the structure.

\subsection{Viscous and capillary pressure drop during pore filling}
The pressure drop during the individual pore filling event is expressed as the sum of the viscous pressure drop of the wetting $\Delta p_w$ and nonwetting $\Delta p_{nw}$ phase and the capillary pressure $\Delta p_c$ drop caused by the meniscus of the fluid front \citep{Guo2025}:

\begin{equation}
    \Delta p= p_1 - p_2= \Delta p_w+\Delta p_{nw} + \Delta p_c
\end{equation}

The pressure drop for the wetting and nonwetting phases can be calculated using the Hagen-Poiseuille law under the assumption of laminar, fully developed flow in cylindrical pores:
\begin{equation}
    \Delta p_w=\frac{8\mu_w L_w v}{r_i^2}
\end{equation}

\begin{equation}
    \Delta p_{nw}=\frac{8\mu_{nw} L_{nw} v}{r_i^2}
\end{equation}

\begin{figure}
    \centering
    \includegraphics[width=0.5\textwidth]{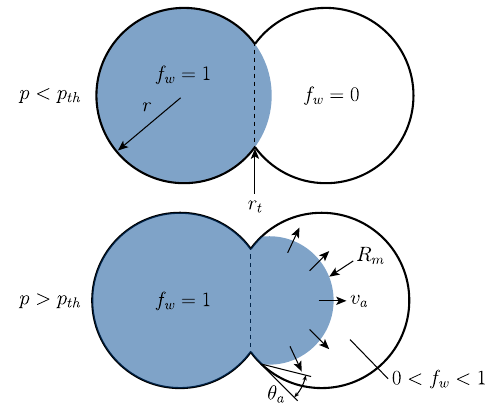}
    \caption{Schematic pore filling mechanism of a network of intersecting hollow spheres.}
    \label{p6:fig_PoreFilling}
\end{figure}

As depicted in Fig.~\ref{p6:fig_PoreFilling}, the investigated porous network consists of connected hollow spheres with the throat being the intersection of two pores of radius $r$ and horizontal distance $l$. Hence, the throat length is $L_t \rightarrow 0$ and therefore offers negligible viscous resistance. As the main viscous dissipation will occur inside the pore body during the rapid acceleration of the liquid and displacement of air, the viscous losses are modeled using the characteristic hydraulic radius of the pore $r$ as a length scale to model the travel distance of the liquid. Within a single capillary tube, the respective wetting and nonwetting phase volume fractions are defined by $f_w=L_w/L$ and $f_{nw}=L_{nw}/L$, allowing a calculation of the effective viscosity $\mu_{\text{eff}}=\mu_wf_w +\mu_{nw}f_{nw}$. 

Assuming that the axial distribution of phases in the pore is proportional to the volumetric filling ratio inside the pore and that the wetting phase enters from one side and progressively displaces the nonwetting phase, one can approximate $f_w$ by:

\begin{equation}
    f_w=V_w/V_p
\end{equation}

Consequently, the effective viscosity $\mu_{\text{eff}}$ is a function of the pore wetting phase volume fraction $f_w$:

\begin{equation}
    \mu_{\text{eff}}(f_w)=f_w\mu_w+(1-f_w)\mu_{nw}
\end{equation}

With that, the viscous pressure loss during the pore filling event is:

\begin{equation}
    \Delta p_v=\frac{16 \mu_{\text{eff}}(f_w)v}{r}
\end{equation}

In the experiment, we imbibe the network filled with the nonwetting liquid with a wetting liquid under the advancing contact angle $\theta_a$. Initially, the meniscus of the fluid front is pinned vertically in the throat of radius $r_t$ before it expands during the jump towards the pore radius $r$. Consequently, $\Delta p_c$ will decrease during the jump. Assuming that the meniscus is a spherical cap of radius $R_m$ spanning the pore, we relate $R_m$ to $f_w$ by using a harmonic interpolation for the meniscus radius:
\begin{equation}
    R_m(f_w)=(\frac{1-f_w}{r_t}+\frac{f_w}{r})^{-1}
\end{equation}

Finally, the capillary pressure drop at the front of the wetting phase meniscus is calculated by:

\begin{equation}
    \Delta p_c(f_w) = \frac{2 \sigma cos\theta_a}{R_m(f_w)}
\end{equation}

\subsection{XMPI experimental setup}\label{p6:XMPI setup}

\begin{figure}[htbp]
    \centering
    \includegraphics[width=0.95\textwidth]{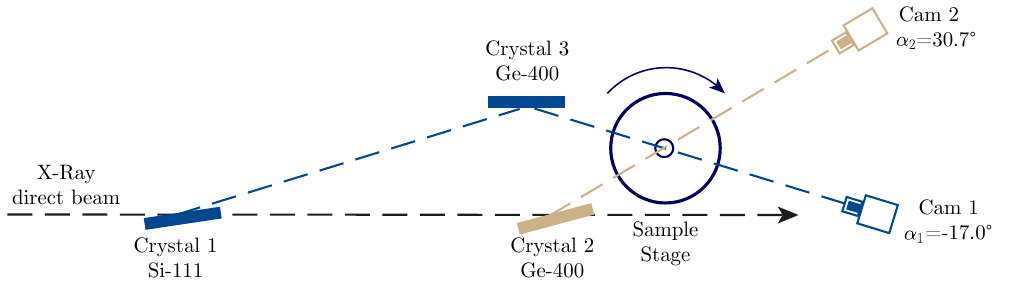}
    \caption{XMPI experimental setup. Crystals 1-3 are used to split the central beam into two beamlets that intersect within the investigated object. Afterwards, the beamlets are captured by X-ray detector systems Cam~1 and Cam~2.}
    \label{p6:fig_xmpi_setup}
\end{figure}

\noindent
The XMPI experiment was performed at the ForMAX beamline \citep{nygard2024} of the MAX~IV synchrotron in Lund, Sweden. We performed the experiments at a photon energy of 16.6 keV, with a direct beam size of 1x1~$\text{mm}^2$ at a photon flux of $10^{15} ph/s/mm^2$.
The experimental setup is described in Fig.~\ref{p6:fig_xmpi_setup}, where two beamlets are established to illuminate the sample simultaneously.
The core of the setup is the beam splitters, inspired by concepts developed in interferometers, split-and-delay lines and specific setups for single-shot imaging~\citep{oberta2013laue,mokso2015simultaneous}. 
The first beamlet was established by splitting the direct beam with a Si-111 crystal and recombining it using a Ge-400 crystal (dark blue line in Fig.~\ref{p6:fig_xmpi_setup}) at -17.0° relative to the direct beam.
The second beamlet was established using a Ge-400 crystal (brown line in Fig.~\ref{p6:fig_xmpi_setup}), resulting in 30.7° with respect to the direct beam.
Both beamlets were properly aligned so that they passed through the same point.
This point is referred to as the intersection point where the sample stage should be positioned.
In this work, the sample stage is continuously rotating during the XMPI experiment, minimizing the difficulty of 4D reconstruction introduced later in Sect.~\ref{p6:Sec_ReconstructionAlgorithm}.
Two beamlets were then detected by two identical indirect X-ray microscopes.
Each X-ray microscope includes a GAGG:Ce scintillator, a 5X magnification lens, and an Andor Zyla 5.5~sCMOS camera.
Each X-ray microscope provided a 50~Hz acquisition rate and a 1.3~µm effective pixel size.

\begin{figure}[htbp]
    \centering
    \includegraphics[width=0.95\textwidth]{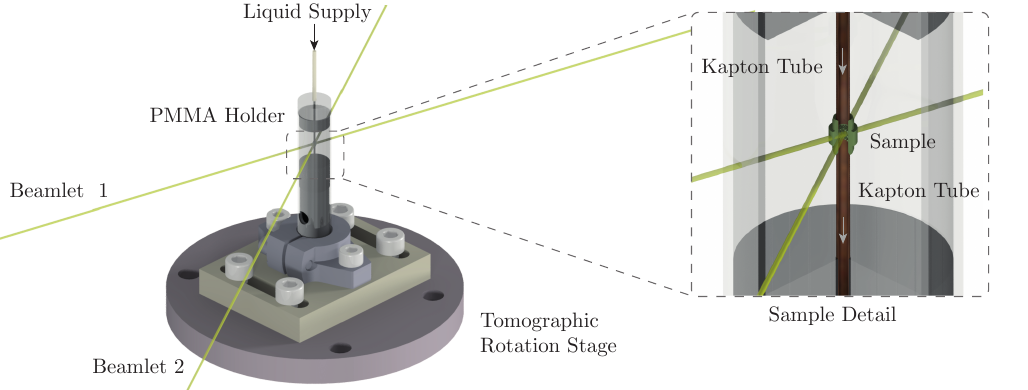}
    \caption{Tomographic rotation stage with PMMA~cylinder and a mounted sample. The beamlets intersect within the sample after transmitting the PMMA holder while a connection to Kapton tubing allows for generating a flow through the sample.}
    \label{p6:fig_Sample Holder}
\end{figure}

The sample itself is the additive-manufactured capillary, which needed to be precisely aligned in a defined position in space while at the same time allowing a liquid flow and rotational movement. Therefore, we are using a setup as sketched in Fig.~\ref{p6:fig_Sample Holder}. A polished PMMA cylinder with a wall thickness of $s$=0.5~mm is used as an X-ray transparent sample holder that is mounted on a tomographic rotation stage, allowing for a slight but continuous rotation. Two PMMA plugs with $d$=0.75~mm holes allow for placing the sample stacked on rigid Kapton tubes (Allectra 312-KAP-TUBE-07-300) within the hollow cylinder, which ensured fixed spatial positioning of the sample in the beam intersection point. Custom slip rings in the PTFE~tubing allowed for a flow of liquid during simultaneous rotation of the setup on the rotation stage. As a result, we can illuminate the flow in the sample with different beamlets while simultaneously obtaining a slight and continuous rotational movement of the samples with constant X-ray attenuation of the PMMA holder.

The flow is driven by a syringe pump (New Era Pump Systems NE-4000) with a 1~ml syringe to create a constant flow of deionized~(DI) water at a rate of $Q=0.5$~ml/h. The vertical flow in the capillary of $2R$=0.7~mm with a bulk velocity of $u_p$=0.36~mm/s can be characterized by the Reynolds number~$Re$ and the capillary number~$Ca$:

\begin{equation}
    Re=\frac{\rho_p\cdot2R\cdot u_p}{\mu_p}=0.25
\end{equation}

\begin{equation}
    Ca=\frac{\nu_p\cdot\rho_p\cdot u_p}{\sigma}=5.0\cdot 10^{-6}
\end{equation}

Consequently, we create a creeping flow regime ($Re<1$, Stokes flow) where inertial forces are small compared to viscous forces and can be neglected. Furthermore, capillary forces dominate over viscous forces, which is typical for microfluidic flows in porous media.

\subsection{Data processing and 4D reconstruction}
\label{p6:Sec_ReconstructionAlgorithm}
In order to utilize the images acquired by both X-ray microscopes presented in Fig.~\ref{p6:fig_xmpi_setup} to reconstruct the dynamics in~4D, implementing proper data processing and reliable 4D reconstruction algorithm is needed. 
The purposes of the data preprocessing are: i) to reduce the noise from the raw data, and ii) to ensure consistency across images from different X-ray microscopes. 
More details of the data processing pipeline can be found in Appendix~\ref{p6:consistency}.

After data processing, X-Hexplane~\citep{hu2025super}, a computationally efficient 4D~Deep-Learning~(DL) reconstruction framework is used to reconstruct the dynamics in~4D.   
Unlike state-of-the-art time-resolved tomography~\citep{garcia2021tomoscopy} performed at synchrotron light sources, which requires a scanning angle of 180 degrees to reconstruct a single time point, our workflow can provide the time-resolved 3D reconstruction at the same frame rate as the 2D~movies acquired by our X-ray microscopes. More details of the 4D reconstruction framework can be found in Appendix~\ref{p6:Hexplane}. 

\subsection{Multiphase Lattice Boltzmann Simulation}
The 4D~pore filling reconstruction was compared to numerical simulations based on the Lattice Boltzmann~(LB) Method, a mesoscopic approach particularly well-suited for simulating multiphase fluid dynamics in complex pore-scale geometries \citep{Ramstad2019} derived directly from synchrotron tomography data. Fundamentally, it is based on solving a discretized Boltzmann equation for fluid particle distributions that stream and collide on a defined lattice space.

Within the underlying study, we are using a lattice type D3Q19, allowing a three-dimensional simulation with $m=19$ discrete velocity vectors $\textbf{c}_i$. The lattice Boltzmann equation describes the distribution dynamics of a fluid particle distribution $f(x,v,t)$ on discrete lattice nodes $\textbf{x}$ by uniting the advection and collision of the fluid distributions $f_i(\textbf{x},t)$ \citep{Ramstad2019}:

\begin{equation}
f_{i}(\mathbf{x} + \mathbf{c}_{i} \Delta t,\, t + \Delta t) - f_{i}(\mathbf{x}, t) = 
\sum_{j} \Lambda_{ij} (f_{j} - f_{j}^{\text{eq}}) + S_{i}(\mathbf{x}, t)
\label{p6:Eqn_LBE}
\end{equation}

While the left-hand side of Eq.~\ref{p6:Eqn_LBE} describes the streaming process that allows fluid distributions to spread with a rate proportional to the velocity vector~$\textbf{c}_i$ to neighboring lattice nodes, the right-hand side consists of a collision term with the scattering matrix~$\Lambda_{ij}$ and a source term~$S_i(\textbf{x},t)$, that allows to apply driving forces or pressure fields. As the scattering matrix~$\Lambda_{ij}$ is difficult to handle directly, a common approach is to substitute $\Lambda_{ij}$ with the nonzero eigenvalue~$\lambda=-\omega$, thereby obtaining a single-relaxation time~(SRT) lattice collision operator, also called Bhatanagar-Gross-Krook~(BGK) collision operator \citep{Bhatnagar1954}. With that, Eq.~\ref{p6:Eqn_LBE} can be simplified:

\begin{equation}
f_{i}(\mathbf{x} + \mathbf{c}_{i} \Delta t,\, t + \Delta t) - f_{i}(\mathbf{x}, t) 
= -\omega \big( f_{i}(\mathbf{x}, t) - f_{i}^{\text{eq}}(\mathbf{x}, t) \big) 
\label{p6:Eqn_SRTSimplification}
\end{equation}

To calculate the relaxation of~$f$ on the right-hand side, the pseudo-equilibrium distribution function~$f^{eq}$ is used, which is a low-velocity Taylor expansion of the Maxwell-Boltzmann distribution \citep{Shan1998}:

\begin{equation}
f_{i}^{\text{eq}} = \rho\, w_{i} \left( 1 + \frac{1}{c_{s}^{2}} \mathbf{c}_{i} \cdot \mathbf{u} + 
\frac{1}{2c_{s}^{4}} (\mathbf{c}_{i} \cdot \mathbf{u})^{2} - 
\frac{1}{2c_{s}^{2}} |\mathbf{u}|^{2} \right)
\end{equation}

For each discrete velocity direction, $w_i$~is the weighting factor of the distribution function. Furthermore, $c_s$~denotes the lattice speed of sound~$c_s=c/\sqrt{3}$. The discrete kinetic model is connected to the macroscopic flow quantities by using the zeroth and first order kinetic moments of the fluid distributions~$f_i(\textbf{x},t)$:

\begin{equation}
\rho(\mathbf{x}, t) = \sum_{i} f_{i}(\mathbf{x}, t)
\end{equation}

\begin{equation}
\rho\, \mathbf{u}(\mathbf{x}, t) = \sum_{i} f_{i}(\mathbf{x}, t)\, \mathbf{c}_{i} 
\end{equation}

With that, macroscopic flow situations can be calculated for Mach~numbers~$Ma=|\textbf{u}|/c_s<0.1$ and small Knudsen~numbers~$Kn=l/L \ll l$.

To simulate the multiphase flow through the porous network, we are using the Shan-Chen model \citep{Shan1993, Shan1994} implemented in a multiphase Lattice-Boltzmann library \citep{Santos2022}, using an existing Lattice-Boltzmann solver~\citep{Latt2021}. The principal idea in the Shen-Chen model is to use the BGK~collision operator and define two separate probability density functions for both fluids according to Eq.~\ref{p6:Eqn_SRTSimplification}. The probability distribution functions for the fluids~$\sigma$ and~$\overline{\sigma}$ interact with each other by a cohesive force that depends on the interparticle force parameter~$G_C$ \citep{Shan1995,Pan2004}:

\begin{equation}
F_{\alpha, \sigma}(\mathbf{x}, t) = -G_{c} \rho_{\sigma}(\mathbf{x}, t) 
\sum_{i} w_{i}\rho_{ \overline{\sigma}}(\mathbf{x} + \mathbf{c}_{i} \Delta t, t) \mathbf{e}_{i}.
\end{equation}

From that, one can derive a correlation for the pressure in the multiphase D3Q19~lattice \citep{Schaap2007, Shan1996}:
\begin{equation}
    p(x)=(\rho_{\sigma}+\rho_{\overline{\sigma}})c_s^2+c_s^2G_c\rho_{\sigma}\rho_{\overline{\sigma}}
\end{equation}

Furthermore, the interactions of the fluid with the solid walls are described using a parameter~$G_{ads}$ \citep{Martys1996}:
\begin{equation}
F_{\text{ads}, \sigma}(\mathbf{x}, t) = -G_{\text{ads}, \sigma} \rho_{\sigma}(\mathbf{x}, t) 
\sum_{i} w_{i} \mathbf{e}_{i} \cdot s(\mathbf{x} + \mathbf{c}_{i} \Delta t, t).
\end{equation}

Based on that, the solid-liquid contact angle~$\theta_{sl}$ is calculated following the model of \cite{Huang2007}:

\begin{equation}
\cos \theta_{\sigma} = 
\frac{G_{\text{ads}, \bar{\sigma}} - G_{\text{ads}, \sigma}}
{G_{c} \, \frac{\rho_{\sigma} - \rho_{d, \bar{\sigma}}}{2}}.
\end{equation}

To compare the simulation results with the experiment, it is crucial to convert the physical quantities in lattice units and vice versa. The lattice viscosity is commonly defined with the relaxation time~$\tau=1/\omega$:
\begin{equation}
    \nu_l=c_s^2(\tau-0.5)\Delta t
\end{equation}

Each step~$\Delta t$ in the lattice scale with resolution~$x_l$ equals a certain timestep~$\Delta t_p$ in the physical scale, which can be calculated using the viscosity ratio of the dynamic viscosity of water~$\mu_w$ and the kinematic lattice viscosity~$\nu_l$:

\begin{equation}
    \Delta t_p=\rho_w\cdot\nu_l\cdot\frac{x_{l}^2}{\mu_w}
\end{equation}

That allows a correlation of the physical flow velocity~$u_p$ and the lattice flow velocity~$u_l$:
\begin{equation}
    u_p=u_l\cdot\frac{\Delta x_l}{\Delta t}
\end{equation}

\begin{figure}[htbp]
    \centering
    \includegraphics[width=0.95\textwidth]{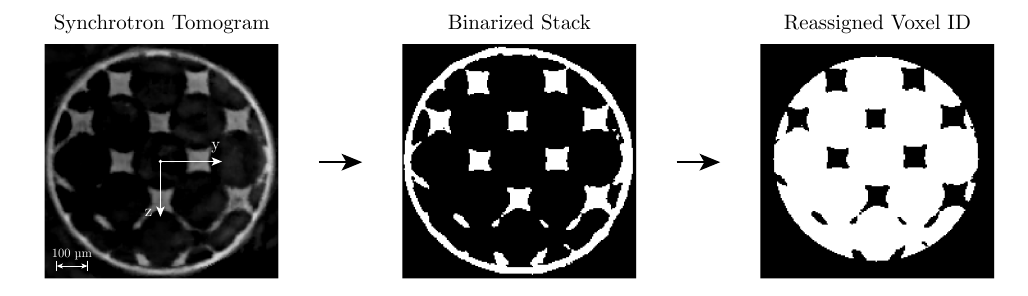}
    \caption{Process of creating a binarized simulation geometry out of the synchrotron tomograph. Process is shown for stack image 34 in the yz-plane.}
    \label{p6:fig_Binarization}
\end{figure}
 
The liquid imbibition is simulated directly on the acquired dry 3D~tomogram, having the advantage that any printing defects and artifacts are captured in the simulation domain. To prepare the recorded tomograph for the flow simulation, a dedicated workflow is necessary, as visible in Fig.~\ref{p6:fig_Binarization}. After reconstructing the greyscale tomograph, it is binarized using the Otsu method \citep{Otsu1979} for thresholding. Furthermore, solid and void voxels are assigned to the correct voxel~ID as a preprocessing step for the simulation. Also, the outer area of the domain is filled with solids and the outer walls are smoothed by applying a circular filter to prevent a destabilized displacement of the fluid front by high surface roughness \citep{Zhou2024}. 10 additional layers are added at the inlet to ensure a stationary flow profile is established before the imbibition process starts.
As we run the simulation based on the acquired tomograph, it would be possible to use a lattice resolution of $x_l$=1.3~µm, rendering a temporal resolution of $\Delta t_s$=71225 steps to equal one frame acquired at 50~Hz rate. To execute the simulation in reasonable timescales \citep{Schaap2007}, we rescale the resolution of the tomogram to $x_l=$5.2~µm, causing $\Delta t_s$=4454~steps to equal one frame in the experiment, which allows reasonable simulation times. 

The boundary conditions for the simulation were chosen to maintain identical flow properties in terms of Reynolds number $Re$ and Capillary Number $Ca$. Furthermore, the static solid-liquid contact angle is set to $\theta_{sl}$=84.9°, based on the slightly hydrophilic behaviour of the cured resin.


\begin{figure}[htbp]
    \centering
    \includegraphics[width=1\textwidth]{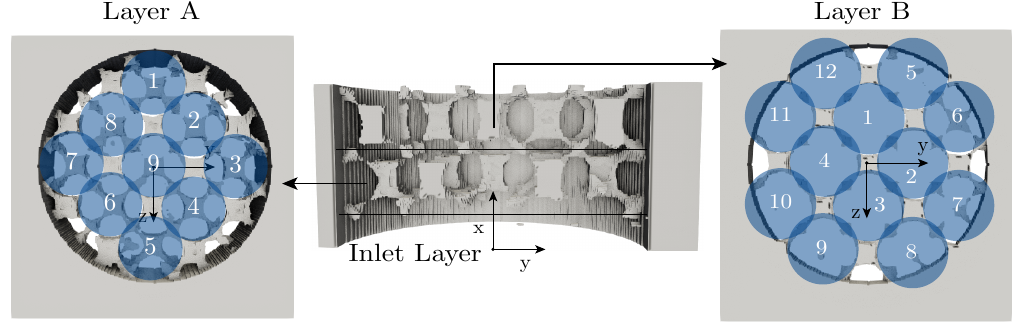}
    \caption{Coordinate system and pore number assigned to the 3D~tomogram of the printed geometry. Layer~A is defined by $20\leq x_{\text{slice}} \leq 31$ while layer~B is defined by $48\leq x_{\text{slice}} \leq 59$.}
    \label{p6:fig_Map}
\end{figure}

\section{Results and Discussion}
\label{p6:Sec_Results}
For comparison of simulation and experiment, each pore was assigned a number as depicted in Fig.~\ref{p6:fig_Map}. This also allows for evaluating the filling ratio of individual pores in the respective layers~A and~B.\\

\subsection{Simulation Results}

Before evaluating the simulations, we evaluated the average values of the dimensionless numbers within the first 500 simulation iterations and determined $Re=0.321$ and $Ca=6.40\cdot10^{-6}$, which are in the same order of magnitude as applied in the experiment.

\begin{figure}[htbp]
    \centering
    \includegraphics[width=\textwidth]{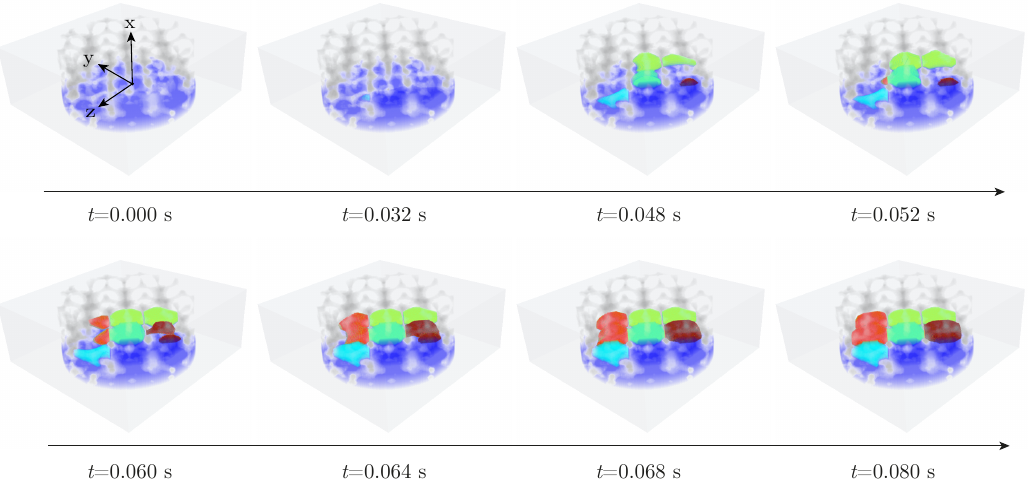}
    \caption{3D simulation results. Time is set to t=0 when fluid enters the simulation domain at $x=0$. The colorscheme of the individual pores is defined by the time of the start of the individual pore filling, see Fig.~\ref{p6:fig_ResultMap} as a reference. The filling starts at pores~A6 and A7, followed by asymmetric progression through others.}
    \label{p6:Fig_3DSimResults}
\end{figure}

Based on the multiphase LB simulation, we investigate the pore filling in 3D for subsequent physical timesteps, as depicted in Fig.~\ref{p6:Fig_3DSimResults}. It is apparent that pore~A6 is nearly filled at $t=0.032$~s, while pore~A7 is completely filled at $t=0.048$~s. Hence, the filling starts from the sides and not from the center of the capillary. At $t=0.060$~s, pores A9, A2 and A1 are partly filled while pores A4, A3 and A8 are still empty. At $t=0.080$~s, these pores are filled and also the following vertical layer is penetrated, starting with pore B1, causing again an asymmetric filling behaviour.

\subsection{Experimental results}

\begin{figure}[htbp]
    \centering
    \includegraphics[width=.5\textwidth]{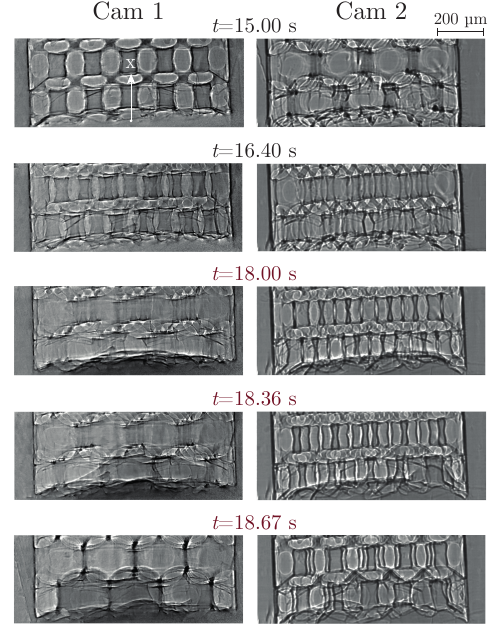}
    \caption{Flat-filed corrected images acquired by XMPI~setup at different times. While stage~1 (marked in black) is present at t=15.00~s and t=16.40~s, liquid is visible from t=18.00~s on, introducing stage~2 (marked in red). The original time of the experiment is used to show stages~1 and~2.}
    \label{p6:fig_xmpi_images}
\end{figure}

Using the experimental setup depicted in Section~\ref{p6:XMPI setup}, we acquired movies from both cameras simultaneously, with the sample stage rotating at a constant speed of 12°/s. The entire movie can be categorized into two stages. During stage~1, no flow enters and only rotation of the sample can be observed. During stage~2, the flow enters so that both the investigated dynamic behaviour and the rotation can be clearly observed.

Although the main scope of this work is to study the dynamic behaviour at stage~2, images at stage~1 still play a crucial role, as they help ensure the self-consistency of the images collected from different cameras and boost the performance of the 4D~reconstruction. Examples of flat-field corrected~\citep{VanNieuwenhove2015} images from both cameras at some typical time points are shown in Fig.~\ref{p6:fig_xmpi_images}.
As indicated in appendix~\ref{p6:consistency}, both cameras provide self-consistent information that is further used for the reconstruction.

\begin{figure}[htbp]
    \centering
    \includegraphics[width=\textwidth]{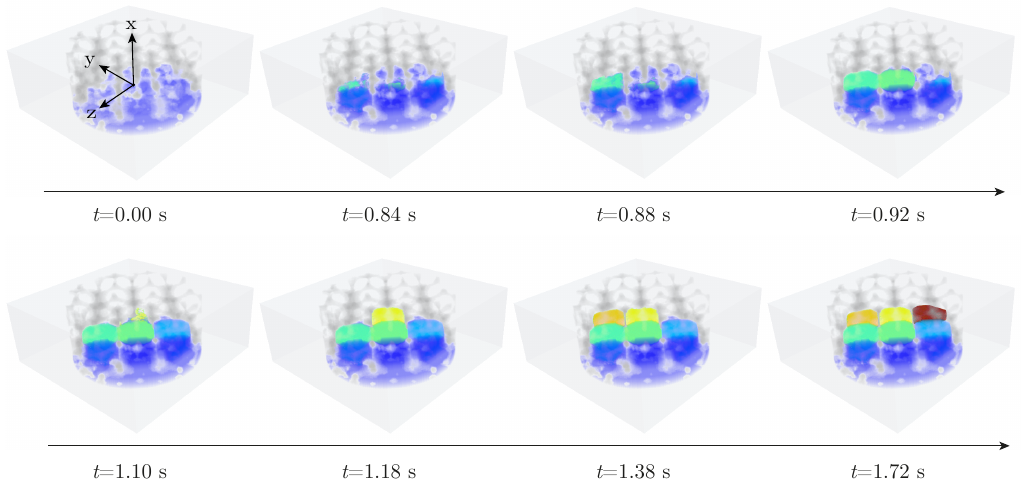}
    \caption{3D reconstruction of pore filling in layer A at selected time points. Time is set to $t = 0$ when fluid first enters the field of view. Filling sequence begins at pores~A6, A7, and A5.}
    \label{p6:fig_Paraview_exp}
\end{figure}

Using the 4D~reconstruction workflow described in Section~\ref{p6:Sec_ReconstructionAlgorithm}, the pore filling dynamics in~3D at different time points for layer~A are shown in Fig.~\ref{p6:fig_Paraview_exp}. It can be seen that the imbibition begins at pores~A6, A7, and A5, which are completely filled at $t=0.84$~s, while the imbibition of pore A8 starts at $t=0.88$~s. The other pores in the central row, A9 and A4, fill at $t=0.92$~s and $t=1.10$~s, respectively. Afterwards, the diagonal row A1, A2 and A3 fills, thereby completing the imbibition of layer~A at $t=1.72$~s.

\subsection{Comparison}

\begin{figure}[htbp]
    \centering
    \includegraphics[width=1\textwidth]{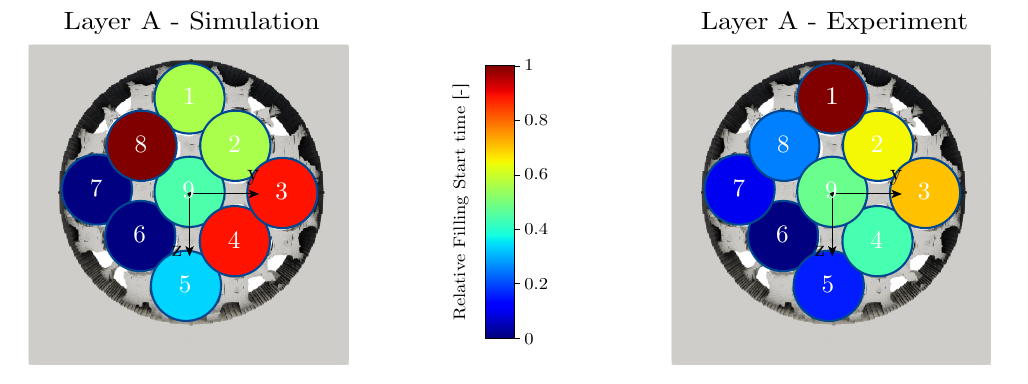}
    \caption{Normalized pore filling order for layer~A in simulation and experiment. Colors indicate relative start time normalized to total layer filling. Simulation: A6 → A7 → A5 → A9 → A1 → A2 → A3 → A4 → A8; Experiment: A6 → A7 → A5 → A8 → A4 → A9 → A2 → A3 → A1.}
    \label{p6:fig_ResultMap}
\end{figure}

Finally, we compare the simulation with the experimental results. Even though comparable Reynolds and Capillary numbers are obtained, the overall filling process is considerably faster in the simulation. While the first layer fills in the simulation within $\Delta t_{\mathrm{sim}}=0.10~\mathrm{s}$, it takes roughly $\Delta t_{\mathrm{exp}}=1.6~\mathrm{s}$ to fill the first layer in the experiment. We attribute these temporal differences primarily to the boundary conditions: In the simulation, a fixed pressure drop is imposed, whereas in the experiment, a constant flow rate is applied. Consequently, in the experiment, the threshold pressure for each pore must be built up from the supplied flow, whereas in the simulation, it may already be exceeded at $t=0$, leading to faster pore filling. While drawbacks in using the pressure boundary condition for simulating Haines jumps are known \citep{Zacharoudiou2018}, the used multiphase model does not allow for prescribing a volumetric flow rate. Also, temporal discrepancies in multiphase simulations and experiments have been reported in~\cite{Yiotis2021} and~\cite{Yan2025}. The observed ~10× difference in filling duration between simulation and experiment is primarily due to the constant flow rate in the experiment, which imposes a supply limitation for the pore filling processes during the imbibition the first pores. This shows the need for reservoir-based designs to decouple supply limitations for future experiments.

To enable a direct comparison of the filling order, we normalize the filling time of the entire layer~A in both cases. Qualitatively, the filling order of pores within layer~A is obtained from the relative time at which filling starts for each pore. This information is color-coded in Fig.~\ref{p6:fig_ResultMap}. 

Both simulation and experiment show sidewise liquid imbibition (pores~A6 and~A7), followed by A5. Divergences occur thereafter: in the simulation, pores~A9, A1, A2 and A3 and fill next, whereas in the experiment the next diagonal row of pores~A8, A4 and A9 fills before A2, A3 and A1. We attribute these differences to the binarization process used to generate the simulation domain from the tomogram. This step removes fine-scale surface roughness and resin residues, altering local curvature radii and thus local threshold pressures $p_{\mathrm{th}}$. The strong influence of the wall microstructure on the simulation results has also been reported by \cite{Bultreys2024}, stating that accurately capturing the contact line dynamics is essential for pore-scale simulations. Additionally, the apparent contact angle at an advancing front is a dynamic quantity that depends on both the static contact angle and the Capillary number~\citep{Yiotis2021}, an effect not captured in our simulation. Nevertheless, the sidewise filling pattern is visible in both cases.

\begin{figure}[htbp]
    \centering
    \includegraphics[width=1\textwidth]{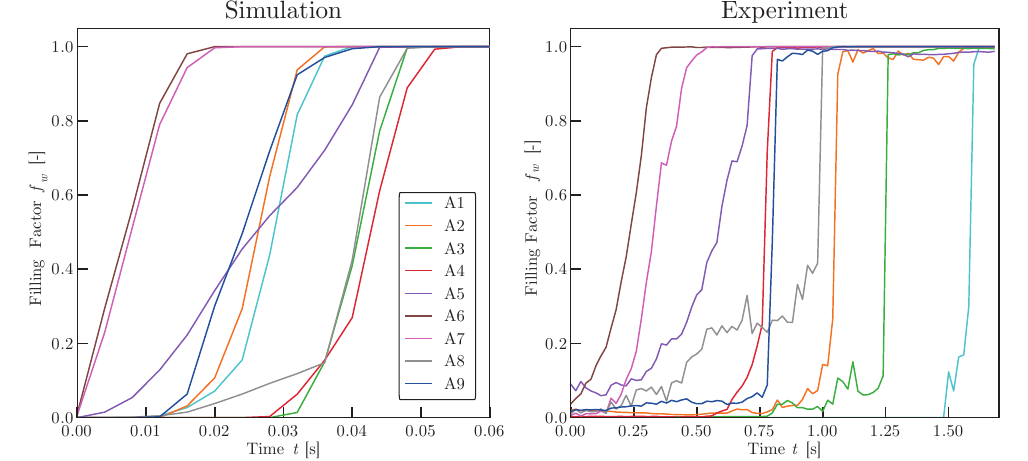}
    \caption{Filling factor $f_w$ vs. normalized time for layer~A. Both simulation and experiment show stepwise pore filling (Haines jumps). Time normalized to first pore filling for comparison.}
    \label{p6:fig_FillingFactor}
\end{figure}

A focus of this study is the analysis of flow instabilities during imbibition of porous networks. We therefore examine the filling factor $f_w=V_w/V_p$ of individual pores as a function of time as depicted in Fig.~\ref{p6:fig_FillingFactor}. For better comparison, we define $t=0$ in both cases at the immediate start of the filling of the first pore. Both simulation and experiment exhibit a pronounced step-wise filling of individual pores, despite the aforementioned timescale differences. Pore~A5 shows a notably slow filling in both cases, which we attribute to printing-induced irregularities present in both the physical geometry and the tomogram.

For quantitative comparison, we define a lower filling threshold $f_{w,\mathrm{lt}}=0.15$ for filling onset and an upper threshold $f_{w,\mathrm{up}}=0.95$ for complete filling. This yields average pore filling durations of $\Delta t_{\mathrm{sim}}=0.0146~\mathrm{s}$ for the simulation and $\Delta t_{\mathrm{exp}}=0.184~\mathrm{s}$ for the experiment, differing by over one order of magnitude. The imaging system records at a temporal resolution of 50~Hz; for several pores (e.g. A2 and A9) filling occurred within only 1-2 frames, as indicated in the overview in Table~\ref{p6:tab_FillingTable}. This suggests that these events may approach the intrinsic Haines jump timescales in the millisecond range. However, the experimental setup imposes a supply limitation: with a pore volume of $V=33.5~\mathrm{nl}$ and a constant inlet flow rate of $0.5~\mathrm{ml/h}$, the minimum time to supply a single pore volume is $t_{\mathrm{min}}\approx 0.24~\mathrm{s}$. This already exceeds the ms-scale bursts reported for idealized systems, indicating that the observed durations are limited by bulk liquid supply rather than intrinsic pore-scale dynamics. The supply path consists of a $d=0.7~\mathrm{mm}$ tube feeding only nine pores in the first layer, so instantaneous flow rates to an individual pore are insufficient to sustain ms-scale jumps. Furthermore, the relatively high advancing contact angle $\theta\approx 84^\circ$ reduces the driving capillary pressure by $\cos\theta\approx 0.14$ compared to a fully wetting liquid, further lowering the possible jump velocity. Consistently, later-filling pores in the sequence show shorter filling durations close to the millisecond scale, as liquid from neighboring pores can contribute to the filling, bypassing the bulk supply bottleneck. This is not visible in the simulation where the filling time does not accelerate at later filled pores.

\renewcommand{\arraystretch}{0.9}
\begin{table}[htbp]
\centering
\small
\setlength{\tabcolsep}{4pt} 
\caption{Relative filling start time, pore filling duration and relative filling velocity for the pores in layer A for the simulation and the experiment.}
\label{p6:tab_FillingTable}
\begin{tabular}{@{}lccc ccc@{}}
\toprule
\multirow{2}{*}{Pore} & \multicolumn{3}{c}{Simulation} & \multicolumn{3}{c}{Experiment} \\ 
\cmidrule(lr){2-4} \cmidrule(lr){5-7}
 & \begin{tabular}[c]{@{}c@{}}Rel. Fill.\\ Start [s]\end{tabular}
 & \begin{tabular}[c]{@{}c@{}}Filling\\ Duration [s]\end{tabular}
 & \begin{tabular}[c]{@{}c@{}}Filling\\ Vel. [mm/s]\end{tabular}
 & \begin{tabular}[c]{@{}c@{}}Rel. Fill.\\ Start [s]\end{tabular}
 & \begin{tabular}[c]{@{}c@{}}Filling\\ Duration [s]\end{tabular}
 & \begin{tabular}[c]{@{}c@{}}Filling\\ Vel. [mm/s]\end{tabular} \\ 
\midrule
A1                                        & 0.417     & 0.012    & 16.7    & 0.959     & 0.060    & 3.3     \\
A2                                        & 0.417     & 0.012    & 16.7    & 0.622     & 0.040    & 5.0     \\
A3                                        & 0.667     & 0.012    & 16.7    & 0.676     & 0.140    & 1.4     \\
A4                                        & 0.667     & 0.016    & 12.5    & 0.419     & 0.060    & 3.3     \\
A5                                        & 0.250     & 0.028    & 7.1     & 0.149     & 0.380    & 0.5     \\
A6                                        & 0.000     & 0.012    & 16.7    & 0.000     & 0.220    & 0.9     \\
A7                                        & 0.000     & 0.016    & 12.5    & 0.095     & 0.220    & 0.9     \\
A8                                        & 0.750     & 0.008    & 25.0    & 0.243     & 0.520    & 0.4     \\
A9                                        & 0.333     & 0.016    & 12.5    & 0.459     & 0.020    & 10.0 \\
\bottomrule
\end{tabular}
\end{table}

\begin{figure}[htbp]
    \centering
    \includegraphics[width=1\textwidth]{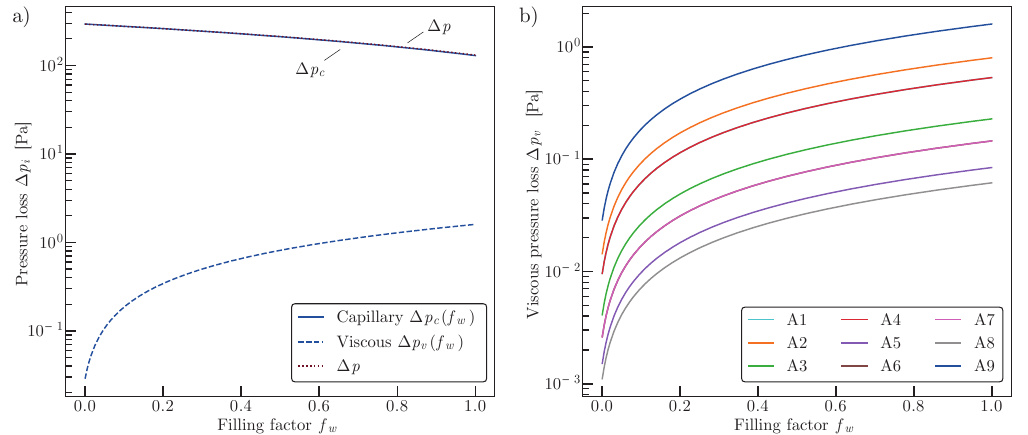}
    \caption{a) Pressure losses as a function of the filling factor $f_w$ for pore~A9. 
    b) Viscous pressure losses as a function of $f_w$ for all pores in layer~A.}
    \label{p6:fig_losses}
\end{figure}

Using the correlations presented in Section~\ref{p6:Sec_Methods}, we calculated the viscous and capillary pressure losses during the vertical filling event of the individual pores in layer~A. As a simplification, we assume that the static solid-liquid contact angle~$\theta_{sl}$ equals the advancing contact angle~$\theta_a$. Fig.~\ref{p6:fig_losses}a shows results for pore~A9, which fills within 20 ms, consistent with Haines jump timescales. The capillary pressure loss, determined from the pore and throat geometry, is identical for all pores and decreases linearly with $f_w$, from $295$~Pa at $f_w=0$ to $129$~Pa at $f_w=1$. The viscous pressure loss increases from 0~Pa to $1.6$~Pa but remains $2$-$4$ orders of magnitude smaller than the capillary term ($\Delta p_v \ll \Delta p_c$). This confirms that the pressure drop during Haines jumps is governed by capillarity, in agreement with~\cite{Guo2025}. The large disparity between $\Delta p_v$ and $\Delta p_c$ persists over the entire filling progression.

The large capillary pressure loss arises from the network geometry and the curvature of the meniscus at the pore-throat interface. Initially, the meniscus is pinned at the throat radius $r_t$, resulting in a high curvature and large $\Delta p_c$. As $f_w$ increases, the curvature radius approaches the pore radius $r$, reducing $\Delta p_c$. The negligible viscous resistance is a direct consequence of the intersecting hollow-sphere geometry with effective throat length $L_t \rightarrow 0$. Without long, narrow ducts, viscous losses arise only from the transient acceleration of the wetting phase and displacement of the nonwetting phase, which are small given the low viscosities of water and air and the short effective flow paths. The gradual increase in $\Delta p_v$ with $f_w$ is explained by the rise in effective viscosity $\mu_{\text{eff}}$ as the wetting phase progressively occupies a larger share of the pore volume. Differences in $\Delta p_v$ of individual pores as depicted in Fig.~\ref{p6:fig_losses}b can be explained by variations in the pore filling velocity, which is constricted by the liquid supply for the first individual pore filling events within our setup. The larger $\Delta p_v/\Delta p_c$ ratios reported in~\cite{Guo2025} can be attributed to the usage of higher-viscosity fluids and longer connecting throats, both of which increase viscous dissipation. In our setup, the low inlet flow rate and supply limitations further reduce observed viscous losses.

\section{Conclusion}
\label{p6:Sec_Conclusion}
In this study, we presented a novel approach to investigate multiphase flow dynamics and instabilities in 4D using synchrotron X-ray multi-projection imaging (XMPI). This technique enables direct visualization of the imbibition of a wetting liquid into a porous medium with unprecedented spatial and temporal resolution, allowing the study of transient pore-scale phenomena such as Haines jumps. Comparison of the experimental results with multiphase flow simulations based on the Shan–Chen Lattice–Boltzmann model revealed that defining the geometry from a greyscale tomogram introduces limitations in simulation accuracy. Additionally, constraints on boundary condition specification and the challenges to resolve the wall microstructure reduce the precision of simulations in capturing pore-scale instabilities, highlighting the need for experimental methods such as XMPI. When combined with additive manufacturing to produce idealized porous networks, XMPI provides a powerful platform for systematically studying pore-scale flow instabilities.

The present work also revealed challenges in applying XMPI to Haines jumps. The limited field of view (1×1~mm²) restricts experiments to networks with only a small number of micropores. Moreover, the low inlet flow rate of 0.5 ml/h imposes supply limitations that limit observable jump velocities. Future experiments should redesign the network to include a reservoir capable of delivering sufficient liquid during a jump event. 
As recently demonstrated, further improvements of the XMPI setup \citep{Asimakopoulo2024, rogalinski2025timeresolved} will enable both projections to provide consistent, high-quality data at temporal resolutions beyond 10 kHz. This will, in turn, support higher-fidelity 4D~reconstructions and open up new possibilities for direct 4D~investigations of Haines jumps on their intrinsic millisecond timescale.

Overall, this work demonstrates that XMPI, when coupled with carefully designed porous networks and appropriate flow control, provides a unique platform for linking pore-scale experiments and simulations, advancing our understanding of multiphase flow instabilities. Further refinement in flow control and simulation fidelity is needed to fully bridge this gap.

\backmatter

\bmhead{Acknowledgements}
The authors would like to thanks R. Westerholz for his support with static CT preparation scans prior to the experiment.
We are grateful to Z. Matej for his support on computing resources and data management at MAX~IV.
We acknowledge MAX~IV Laboratory for the beamtime on ForMAX under Proposal 20240748. 
Research conducted at MAX~IV, a Swedish national user facility, is supported by the Swedish Research Council under contract 2018-07152, the Swedish Governmental Agency for Innovation Systems under contract 2018-04969 and Formas under contract 2019-02496.
This work received funding and support from the ERC-2020-STG, 3DX-FLASH (948426).

\bmhead{Author contribution}
Conceptualization:~[PW, TR, LDS, AG], Methodology:~[PW, PVP, EMA, ZY, ZH, YZ, JKR]; Data Curation~[ZY, PVP, PW], Formal Analysis:~[PW, ZY], Resources:~[KN, ZY, JKR, ZH, YZ, EMA, PVP], Investigation:~[PW, ZY, ZH, JT, EO, SD, JKR, YZ, EMA, PVP], Visualization:~[PW, ZY], Supervision:~[LDS, TR, PVP], Writing~-~original~draft:~[PW, ZY], Writing~-~reviewing~\&~editing:~All~authors

\bmhead{Data availability}
The data that support the findings of this study are available upon reasonable request from the authors.

\section*{Declarations}

\bmhead{Conflict of interest}
The authors declare that they have no competing interests.

\bmhead{Ethical approval}
Not applicable.
\newpage

\begin{appendices}
\label{p6:sec_appendix}

\section{XMPI data preprocessing}
\label{p6:consistency}
Prior to implementing the 4D~reconstruction workflow using the images from Cam~1 and Cam~2, it is crucial to preprocess the images to reduce the noise and ensure consistency across the two cameras. 
Such preprocessing can be summarized as three steps.

\begin{figure}[tb]
    \centering
    \includegraphics[width=0.95\textwidth]{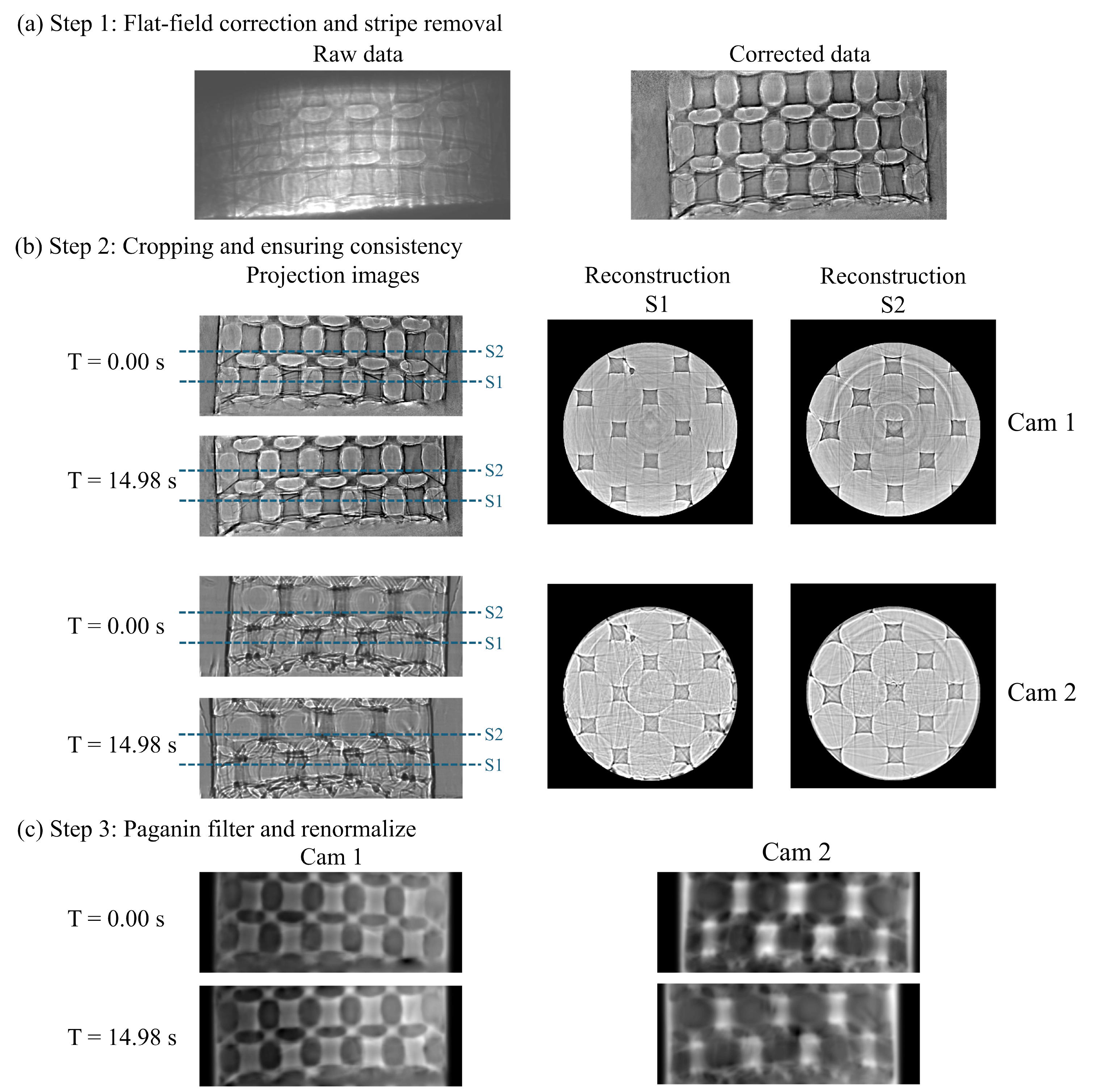}
    \caption{Three steps of the data preprocessing. (a) shows an example of the raw image and the processed image after flat-field correction and stripe removal; (b) shows projection images and reconstructed slices using Cam~1 and Cam~2 at stage~1 (before the flow enters) of the acquired movie; (c) shows examples of the images sent into the reconstruction workflow using X-Hexplane.}
    \label{p6:fig_img_consistency}
\end{figure}

The first step is to apply flat-field correction~\citep{VanNieuwenhove2015} together with the stripe removal technique~\citep{vo2018superior} to reduce background noise. 
An example is given in Fig.~\ref{p6:fig_img_consistency}a.

The second step is to crop the images properly so that both cameras provide consistent information. 
To validate the effect of such cropping, we used noise-reduced images of Cam~1 and Cam~2 from stage~1 (before the flow enters) of the acquired movie, covering a 180-degree rotation and compared the reconstructed slices from Cam~1 and Cam~2, respectively, using the Gridrec~\citep{marone2012regridding} algorithm provided in Tomopy~\citep{gursoy2014tomopy}, an open-source Python package for tomographic data processing and image reconstruction. 
As shown in the right-hand side of Fig.~\ref{p6:fig_img_consistency}b, both Cam~1 and Cam~2 result in similar reconstructed slices, indicating that the cropping provided in the left-hand side of Fig.~\ref{p6:fig_img_consistency} is appropriate.

The final step is to apply the method proposed in \cite{paganin2002simultaneous} as a filter to reduce the edge enhancement effect and then renormalize the images based on the Radon transform property.
Namely, the sum of pixel values in each projection remains constant for each time point.
Processed images after the final step are used as the ground truth projections in the 4D~reconstruction workflow using X-Hexplane, as discussed in Sect.~\ref{p6:Sec_ReconstructionAlgorithm}.  
Examples of processed images after the final step are given in Fig.~\ref{p6:fig_img_consistency}c.

\section{4D reconstruction using X-Hexplane}
\label{p6:Hexplane}
X-Hexplane~\citep{hu2025super} is a powerful tool suitable for reconstructing rotating samples and can be tailored to the XMPI experimental setup presented in this work.
Adapted from Hexplane~\citep{cao2023hexplane}, which is designed for visible light setups, X-Hexplane features i) implementing the physics of X-ray propagation under projection approximation, ii) using a tensorial representation of the dynamics in~4D, and iii) sharing the features over space and time to address the reconstruction from ultra-sparse views provided by the XMPI setup. 

\begin{figure}[tb]
    \centering\includegraphics[width=0.95\textwidth]{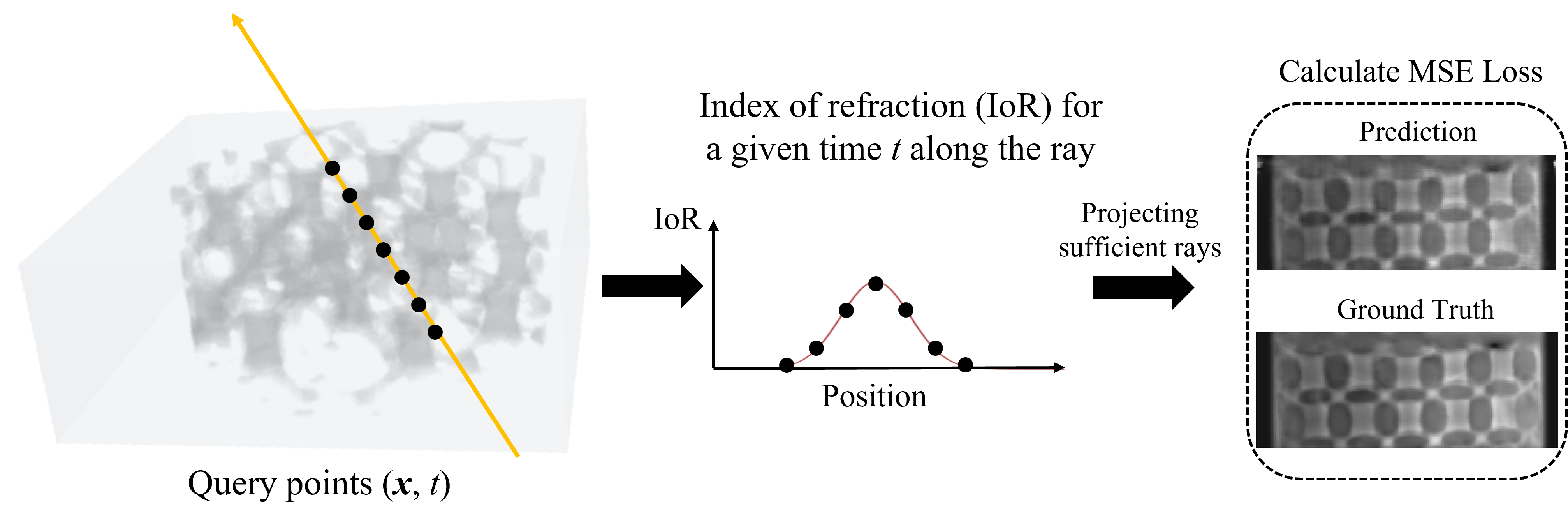}
    \caption{Optimization process of X-Hexplane. 2D projections from a given angle and time point can be rendered using the index of refraction (IoR) along the ray direction. The parameters of X-Hexplane are optimized based on the Mean Squared Error (MSE) loss between the rendered results and the projection images provided by the XMPI setup. Figure adapted from Ref.~\cite{hu2025super}.}
\label{p6:fig_Hexplane}
\end{figure}

The core of X-Hexplane is to project each sampled point $(x, y, z, t)$ onto six feature planes ((X, Y), (X, Z), (X, T), (Y, Z), (Y, T), (Z, T)) to form six corresponding feature vectors. These feature vectors are fused and then fed into the multilayer perceptron (MLP) to generate the index of refraction $n(\boldsymbol{x},t)$ at the specific spacetime point.  
Using $n(\boldsymbol{x},t)$, one can generate projections at any angle at any time point by integrating along the X-ray propagation direction using the projection approximation.
The Mean Squared Error (MSE) between the generated projections and the ground truth projections provided by the XMPI setup is then calculated to optimize the parameters of X-Hexplane, as shown in Fig.~\ref{p6:fig_Hexplane}.

In this work, X-Hexplane was implemented using Python 3.9.18 and PyTorch 1.12.1. The optimization was performed on an NVIDIA V100 GPU with 40GB of RAM. Benefiting from the efficiency of X-Hexplane, it took about 40 minutes to optimize 100000 epochs. 

\end{appendices}

\clearpage
\newpage


\bibliography{cas-refs}

\end{document}